\documentclass{aa}
\usepackage{graphics}
\usepackage{epsfig}

\newcommand{\nh}{N$_{\rm H}$}
\newcommand{\exo}{EXO\,0748$-$676}

\begin{document}                                          
\thesaurus{08.02.2, 13.25.5, 02.01.2, 08.09.2 \exo/UY Vol}
\title{The eclipsing bursting X-ray binary \exo\ revisited by 
XMM-Newton}
\author{J.M. Bonnet-Bidaud\inst{1} 
\and F. Haberl\inst{2}
\and P. Ferrando\inst{1}
\and P.J. Bennie\inst{3}
\and E. Kendziorra\inst{4}} 
\offprints{J.M. Bonnet-Bidaud}
\mail{bobi@discovery.saclay.cea.fr}
\institute
{Service d'Astrophysique, DSM/DAPNIA/SAp, CE Saclay, 
F-91191 Gif sur Yvette Cedex, France
\and
Max-Planck-Institut f\"ur Extraterrestrische Physik, Giessenbachstrasse, 
85748 Garching, Germany
\and Department of Physics \& Astronomy, University of Leicester, 
University Road, Leicester LE1 7RH, UK
\and
Institut f\"ur Astronomie und Astrophysik - Astronomie, University of T\"ubingen, 
Waldh\"auser Strasse 64, D-72076 T\"ubingen, Germany}
\date{Received date: ; accepted date: }
\maketitle

\thesaurus{08.09.2 \exo/UY Vol; 08.02.2; 13.25.5; 02.01.2}

\begin{abstract}
The bright eclipsing and bursting low-mass X-ray binary \exo\
has been observed at several occasions by XMM-Newton during the 
initial calibration and performance verification (CAL/PV) phase.
We present here the results obtained from observations with the EPIC 
cameras.
Apart from several type-I X-ray bursts, the source shows a high degree of variability 
with the presence of soft flares.
The wide energy coverage and high sensitivity of XMM-Newton allows for the first
time a detailed description of the spectral variability.\\
The source is found to be the superposition of a central ($\sim$~2$\times 10^8$ cm) 
Comptonized emission, most probably a corona surrounding the inner edge of an accretion 
disk, associated with a more extended ($\sim$~3$\times 10^{10}$ cm) thermal halo 
at a typical temperature of $\sim$~0.6 keV with an indication of non-solar abundances.
Most of the variations of the source can be accounted for by a variable absorption
affecting only the central comptonized component and 
reaching up to \nh\ $\sim$~1.3$\times 10^{23}$ cm$^{-2}$. 
The characteristics of the surrounding halo are found compatible with an irradiated atmosphere
of an accretion disc which intercepts the central emission due to the system high inclination. 
\end{abstract} 

\section{Introduction}
The X-ray binary \exo\ is unique in showing all types of
variability commonly seen in different low-mass X-ray binaries. The
source was discovered in outburst by the EXOSAT satellite. The
observations revealed the existence of sharp eclipses, intensity dips as
well as repetitive type I bursts which were used to infer the presence
of a neutron star as the compact object and to estimate the distance to 
$\sim$~10 kpc (Parmar et al.\ 1986 hereafter PA86, Gottwald et al.\ 1986).
More recent observations with the Rossi X-ray Timing Explorer (RXTE)
have also shown that the source exhibits variable ($\sim$~Hz and $\sim$~kHz)
quasi-periodic oscillations (Homan et al.\ 1999, Homan \& van der Klis
2000). The length of the orbital period (3.82 hr), combined with the
duration of the eclipses, allows the geometry of the system to be
constrained with a high $\sim$~75$^{\circ}$ inclination and a
$\sim$~0.5 M$_{\sun}$ companion. The presence of residual flux during
eclipses and of dips in the orbital light curve makes \exo\
typical of the ADC (accretion disc corona) sources in which the central
compact source is surrounded by an asymmetric disc with the presence of a
bulge in the outer edge and an extended central hot region (White 
et al.\ 1982).
This complex geometry has prevented so far the different components to be isolated 
unambiguously and contradictory interpretations of the spectral variability
have been put forward.  From EXOSAT data, Parmar et al.\ (1986) concluded
to the presence of a cut-off power-law emission whose part was strongly
absorbed. However from ASCA observations, Church et al.\ (1998) described the
spectrum as the superposition of a point source ($\sim$~2 keV) blackbody from the
neutron star and a partially covered Comptonized emission, originating 
from the ADC.\\
The very rich variability of \exo\ makes it an ideal target for
XMM-Newton. The high throughput and large energy coverage of the EPIC cameras
provide for the first time a detailed picture of the system.
We present here the results obtained with the EPIC MOS and PN cameras.
More results on the RGS Grating observations of the source can be found 
in Cottam et al.\ (2001).

\section{Observations and data reduction}

\exo\ was observed by XMM-Newton at four occasions during the
CAL/PV phase, in revolutions 50, 55, 59 and 67 of the satellite. Details
on the XMM-Newton mission and on the EPIC MOS and PN cameras can be
found in Jansen et al.\ (2001), Turner et al.\ (2001) and Str\"uder et al.\
(2001).

\begin{table}
\caption[ ]{Log of EPIC observations}
\begin{flushleft}
\begin{tabular}{llrrr}
\hline\noalign{\smallskip}
\multicolumn{1}{c}{Instrument } & \multicolumn{1}{c}{Mode$^{1}$/Filter}   
& \multicolumn{1}{c}{SR$^{2}$} &  \multicolumn{1}{c}{Start time } 
& \multicolumn{1}{c}{T$_{\rm exp}$} \\
 & & & \multicolumn{1}{c}{(2000, UT)} & \multicolumn{1}{c}{(ks)}\\
\noalign{\smallskip}
\hline\noalign{\smallskip}
\noalign{\smallskip}
MOS1+2   & FF/Thin    & 59 & Apr. 04, 17:18 & 20.0 \\
MOS1+2   & FF/Medium  & 67 & Apr. 21, 04:11 & 16.9 \\
PN       & FF/Medium  & 67 & Apr. 21, 03:59 & 18.1 \\
PN       & LW/Medium  & 67 & Apr. 21, 09:50 &  3.2 \\
PN       & SW/Medium  & 67 & Apr. 21, 15:22 & 18.1 \\
\noalign{\smallskip}
\hline\noalign{\smallskip}
\end{tabular}
\end{flushleft}
(1) FF Full Frame, LW Large and SW small window mode\\
(2) Satellite revolution\\
\end{table}

Scientific data from EPIC MOS were secured in revolutions 59 and 67 (see Table 1).
The data, obtained with a 2.6 s time resolution, were processed with the
standard analysis of the XMM Science Analysis System. This involves the removal
of bad or hot pixels, and of the electronic noise. For the MOS,
X-ray events with pattern 0 to 12 (similar to grades 0-4 in ASCA) were
selected. The gain was adjusted with reference to the closest observation
in the Closed Calibration position. 
The source events were selected by applying a mask of radius
1.5\arcmin\ centered on the source which corresponds to $\sim$~98\% of the energy
of the typical point spread function of the telescope (Gondoin et al.\ 2000).
The source is bright and some effects of the pile-up of events might be expected
in the FF mode.
The average counting rate in each MOS camera, except for the bursts, was found to be between 
(7-10) event/reading frame which is at the threshold for significant pile-up
(see Fig.~6 in Ballet 1999). The pile-up rate is expected to be less than $\sim$~0.8\%
so that a moderate effect should affect the most central part of the image.

EPIC PN data are currently available from revolution 67 of XMM-Newton
in three different modes (Table 1). 
For light curves we selected the valid event-patterns (single,
double, triple and quadruple; pattern 0-12) while for the spectral
analysis only single-events (pattern 0) were used for which an advanced
spectral response matrix exists. To avoid pile-up effects only the data
from the fastest read-out mode (the SW) was used for the spectral
analysis.

\section{The varying light curve of \exo}

\subsection{The EPIC MOS observations}

Light curves were produced by selecting the source events in two
different soft S (0.5 - 2 keV) and hard H (2 - 10 keV) energy ranges and
adding the counts from the two MOS cameras. The background rate is found 
negligible with mean values
of $\sim$~0.04 and $\sim$~0.08 counts s$^{-1}$, respectively in the soft
and hard bands, illustrating the very clean images obtained by
XMM-Newton. Figures 1 and 2 show the extreme
variability of the source for the two observations. In each observation,
an eclipse is clearly seen in the hard band as well as several type I
bursts with counting rates reaching up to 25 counts s$^{-1}$.

\begin{figure*}
\resizebox{12cm}{!}{\includegraphics{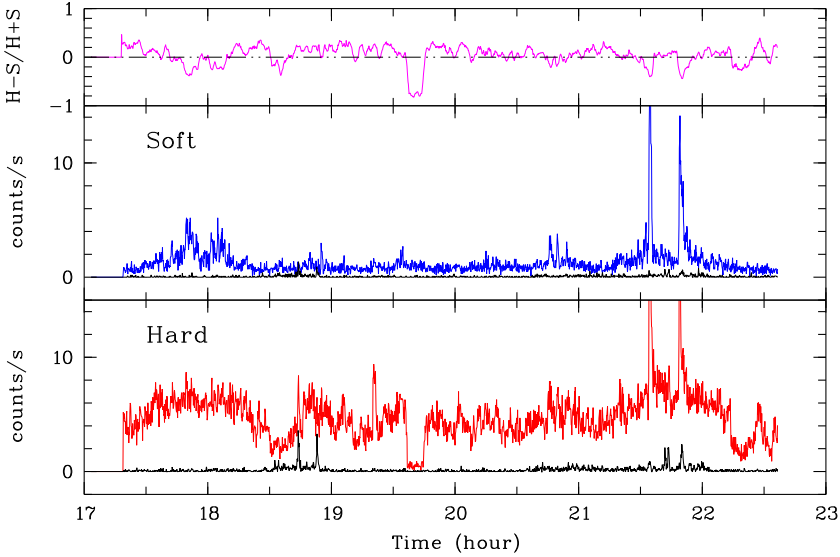}}
\hfill
\parbox[b]{55mm}{
\caption[ ]{The MOS hard H (2 - 10 keV) (bottom) and soft S (0.5 - 2 keV) (middle)
light curves of \exo\ as observed by XMM-Newton on 2000 April 4. 
Time in hours from 2000 April 4, 0:0 UT. The curves
are source counts obtained from the two MOS cameras and binned to 10 s.
In each case, the underlying curve is the background 
rate accumulated in a region of the image free of sources.
The hardness ratio computed as (H-S)/H+S) is shown on the top, where the 
curve have been smoothed by a 5-points average. Note the near total eclipse at 
$\sim$~19:40 UT and the two dips at $\sim$~18:30 UT and $\sim$~22:20 UT. Burst peak counts
are cut in the figure}}
\label{fig-moslc1}
\end{figure*}

\begin{figure*}
\resizebox{12cm}{!}{\includegraphics{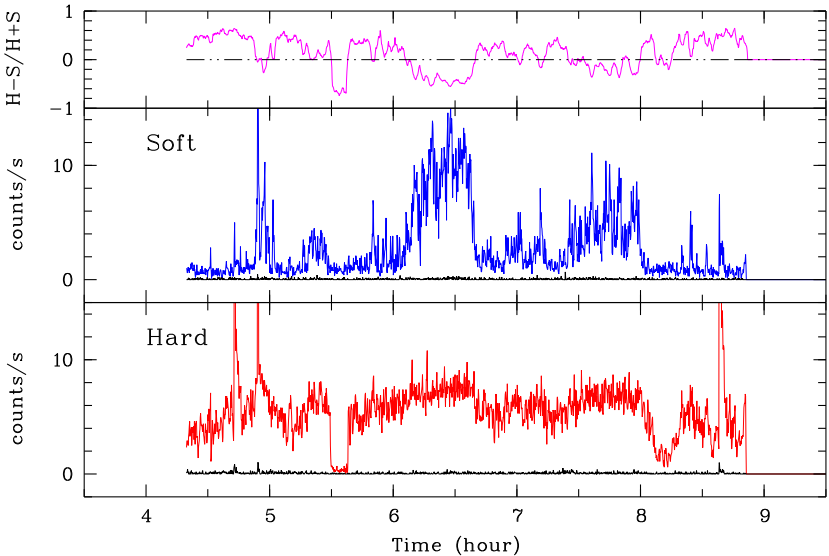}}
\hfill
\parbox[b]{55mm}{
\caption[ ]{The MOS light curves for 2000 April 21, with the same scale as Fig. 1.
Time in hours from 2000 April 21, 0:0 UT.
 Note the two significant soft flares around 06:30~UT and 07:40~UT}}
\label{fig-moslc2}
\end{figure*}

The characteristics of the two eclipses are similar to those reported from 
EXOSAT
(PA86) and RXTE (Hertz et al.\ 1997).
The present lack of exact absolute timing prevents a precise comparison with 
the source ephemeris. However, the use of the provisional start times listed in Table 1
gives observed  mid eclipses times which fall within 17 s (April 4) and 38 s 
(April 21) of the predicted times using the linear ephemeris of Hertz et al.\ (1997). 
More tight constraints on the period changes
will be provided when precise absolute timings will be available.
Due to the very low background rates, the residual counts in eclipse can be 
measured for the first time very accurately. In the hard band, the residual
counts are (0.50$\pm$0.20) counts s$^{-1}$ and (0.37$\pm$0.18) counts s$^{-1}$ 
on April 4 and 21 respectively, which
corresponds to (10$\pm$4)\% and (7$\pm$3)\% of the corresponding mean flux
outside eclipse.
Significantly, this ratio, in the time interval corresponding to the hard eclipse, 
is (87$\pm$25)\% and (32$\pm$10)\% in the soft band.
The behaviour of the source is different in the two observations 
as seen in Fig.\ 1 and 2.

On 4 April 2000, the variability is mainly confined to the hard band with the 
presence of two typical dips, around phase $\sim$~0.7 of the 3.82 hr cycle 
(where phase 0 is defined as the mid-eclipse time).
There is no evidence in the soft band of the eclipse nor of the dips.
The hardness ratio, computed as (H-S)/(H+S), is also shown at the top of 
Fig.\ 1 and 2. A significant softening is seen during eclipse, bursts and dips.
On 21 April 2000, the source appears in a quite different and more active state.
An intense flaring activity is seen in the soft band with 
two major increases of flux arising around phase 0.2 and 0.6. 
The hardness ratio is marked by a significant softening during these 
flares.

A similar activity around the same phases has been noted from 
previous ASCA observations  though above 1 keV  and with somewhat different
energy bands (Church et al.\ 1998). The lack of information 
at lower energy has made the interpretation uneasy in this case. 
The very wide spectral coverage of EPIC allows here a more 
straightforward separation between the lowest energies 
(below 2 keV) where the absorption effects are dominant and 
the higher ones (above 2 keV) where they are negligible (see below). 
We note also that our definition of the hardness ratio is quite different 
to that adopted in the 1 - 10 keV range for the EXOSAT observations
(PA86).

\subsection{The EPIC PN observations}

The EPIC PN observations of \exo\ in FF and SW modes on April 21 
(see Table 1) were performed 
accidentally at almost identical binary phases. The light curves are 
qualitatively identical showing marked features like flaring activity 
and dips at similar phases. Since during the FF observation also the 
MOS cameras observed \exo\ (Fig.\ 2) we show only the 
EPIC PN light curves of the SW observation in Fig.\ 3.

\begin{figure*}
\resizebox{12cm}{!}{\includegraphics{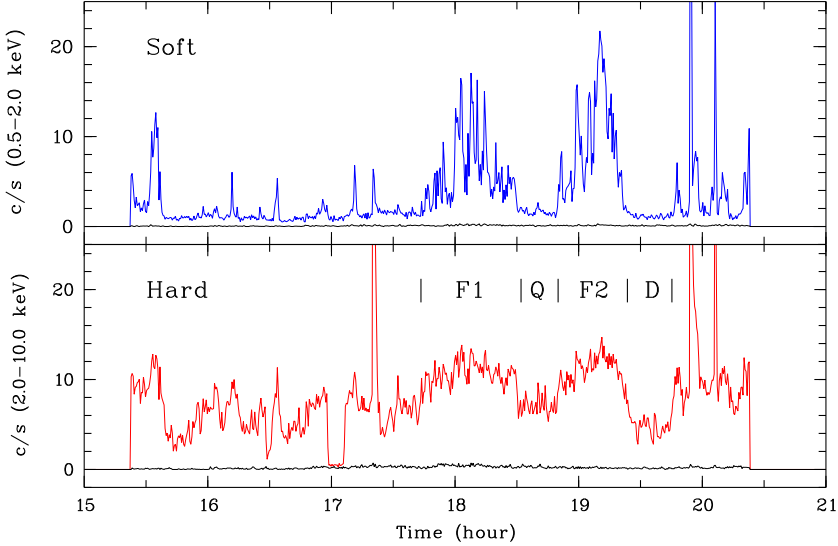}}
\hfill
\parbox[b]{55mm}{
\caption[]{
EPIC PN light curves of \exo\ from the small window mode observation on
2000 April 21 in the 0.5 - 2.0 keV (top) and 2.0 - 10.0 keV energy
bands. Data binning is 30 s and time in hours from 2000 April 21, 0:0 UT.
The background contribution is between 0.05 and 0.2 counts s$^{-1}$ 
in the soft band and 0.05 - 0.5 counts s$^{-1}$ in the hard
band. Note the similarity of the light curves during different binary
orbits (see Fig.\ 2), in particular the stable phasing of the
flaring events
}}
\label{fig-pnlc}
\end{figure*}

\section{Spectral variability}

Source spectra were accumulated from the EPIC PN data, using a circular
extraction region of radius 40\arcsec\ around the peak surface
brightness. Background spectra were created in adjacent regions of the
same size. In view of the large variability of the source four
representative spectra were chosen by selecting events in time intervals
corresponding to the first and second soft flaring episodes (F1 and F2),
the between-flare or quiescent interval (Q) and the dip part (D) of the
EPIC PN SW observation on April 21 (see Fig.\ 3). The count rate during this
observation was well below the pile-up limit of the SW mode of $\sim$~100
counts s$^{-1}$. Spectra were binned to obtain at least 20 counts per
bin. The most recent response matrices (version Sept. 2000) were
obtained from the hardware groups and the spectra analyzed using the
XSPEC package.

Descriptions with a power-law and either a
blackbody or a multi-temperature blackbody accretion disc model, as
suggested from previous observations, were found inadequate (see
Table 2 where reduced $\chi^2$ values are given for various spectral models). 
In particular none of the models represents the spectrum 
around 0.7 keV where a steep decline due to an emission line feature
can be seen. This line can be attributed to the O-VIII ion.
To model the soft part of the spectra below 2 keV where more lines
are visible we therefore tried
thermal emission models, keeping in mind that the models available in
XSPEC were developed for a thin plasma in collisional equilibrium, 
while the plasma which likely exists in \exo\ is photo-ionized
by hard X-ray emission from a central source (see below). 

Best fits in the range 0.15 - 10 keV were obtained with a combination of a
power-law (PL) and a Raymond-Smith thermal component (RS), each
attenuated by individual photo-electric absorption (Morrison \& McCammon
1983). Elemental abundances were treated relative to solar (Anders \& 
Grevesse 1989). It was checked that the fits with the Mewe et al (1985)
instead of the RS thermal model gives fully consistent results including 
for the abundances.

The four representative spectra were jointly fitted by the same model. 
In the power-law
component, intensity and index were each kept as single free parameter
tied together for all four spectra. Only the column density attenuating
this component was allowed to vary individually. Similarly in the
thermal component, the temperature and the elemental abundances were
tied, but intensity and absorption allowed to vary separately. 
The abundances of Si and higher Z elements were fixed to zero
as is suggested by the RGS spectrum  where no 
emission lines of these elements are found (Cottam et al.\ 2001).
The other abundances were allowed to vary freely.
The best fit RS model results in non-solar abundances of N (10$\pm$2), 
O (0.6$\pm$0.1), Ne (0.02$\pm$0.02) and Mg (0.31$\pm$0.23).
The Mg line was only seen in the spectrum between flares. 
Table 3 summarizes the best fit continuum parameters for the different 
intensity levels of the source. All quoted errors were
determined at $\Delta\chi^2$ of +2.7 around the $\chi^2$ minimum. 
The spectra are shown in Fig. 4 with 
superimposed best fit spectral models.
Residual features around emission lines seen below 1 keV are caused by 
calibration problems of the charge transfer losses in SW mode. In the 
current version of charge transfer loss correction in SW the event 
energies are over-corrected by about 10 - 20 eV at energies below 1 keV, 
i.e. the data points in the spectra are shifted to the right. This 
causes the typical residual pattern seen in the bottom panel of
Fig.\ 4 where data and model
emission lines are shifted slightly against each other in energy.
This should affect the derived abundances only slightly and will not
affect our conclusions. Other residuals near 2.2 keV are caused by problems
in the effective mirror areas around the Au edge.

\begin{table}
\caption[ ]{Model fits to the flare spectrum (F1)}
\begin{flushleft}
\begin{tabular}{lr}
\hline\noalign{\smallskip}
\multicolumn{1}{c}{Model$^1$} & \multicolumn{1}{c}{$\chi^{2}_{\rm r}$/dof} \\
\hline\noalign{\smallskip}
powerlaw*wabs             &   6.66/1071\\
(powerlaw+bbody)*wabs     &   2.10/1069\\
powerlaw*wabs+bbody*wabs  &   1.57/1068\\
powerlaw*wabs+diskbb*wabs &   1.62/1068\\
\noalign{\smallskip}
\hline\noalign{\smallskip}
\noalign{\smallskip}
\end{tabular}
\end{flushleft}
(1) Model components as defined in XSPEC
\end{table}

\begin{table*}
\caption[ ]{Best-fitting spectral parameters}
\begin{flushleft}
\begin{tabular}{lrrrrrrr}
\hline\noalign{\smallskip}
\multicolumn{1}{c}{ } & \multicolumn{3}{c}{Power-law}  
& \multicolumn{3}{c}{Thermal component} &  \multicolumn{1}{c}{ }\\
\multicolumn{1}{c}{ } & \multicolumn{1}{c}{A$_{\rm PL}$ (*)} & \multicolumn{1}{c}{$\alpha$ (*)} 
& \multicolumn{1}{c}{\nh} & \multicolumn{1}{c}{A$_{\rm RS}$} 
& \multicolumn{1}{c}{kT (keV) (*)}    
& \multicolumn{1}{c}{\nh} & \multicolumn{1}{c}{$\chi^{2}_{\rm r}$/dof} \\
\noalign{\smallskip}
\hline\noalign{\smallskip}
\noalign{\smallskip}
Flare 1   & 0.0230 (12) & 1.35 (3) & 2.25 (11)  & 0.021 (3)  & 0.64 (5) & 0.041 (6)  & 1.27/2467\\
Quiescent &             &          & 6.43 (28)  & 0.009 (2)  &          & 0.145 (26) & \\
Flare 2   &             &          & 1.48 (10)  & 0.033 (4)  &          & 0.044 (6)  & \\
Dip       &             &          & 13.1 (6)   & 0.011 (2)  &          & 0.22  (3)  & \\
\noalign{\smallskip}
\hline\noalign{\smallskip}
\noalign{\smallskip}
\multicolumn{8}{l}{(*) single parameter in the joint fit to all four spectra}\\
\multicolumn{8}{l}{A$_{\rm PL}$ in photons cm$^{-2}$ s$^{-1}$ keV$^{-1}$ at 1 keV, 
                   $\alpha$ photon index, \nh\ in units of 10$^{22}$ cm$^{-2}$}\\
\multicolumn{8}{l}{A$_{\rm RS}$ emission measure in units of 10$^{-14}$/(4$\pi$d$^2$) $\int$n$_{\rm e}$n$_{\rm H}$dV,}\\
\multicolumn{8}{l}{where d  is the distance to the source (cm), n$_{\rm e}$ and n$_{\rm H}$ are the electron
                   and hydrogen densities (cm$^{-3}$)}\\
\end{tabular}
\end{flushleft}
\end{table*}

The modulation of the source is mainly due to a strong PL absorption varying from
13.0$\times 10^{22}$ cm$^{-2}$ in the dip to $\sim$~6$\times 10^{22}$ cm$^{-2}$ in quiescence
and down to (1.5 - 2.2)$\times 10^{22}$ cm$^{-2}$ during flares. In the same time, the soft
RS component is affected by a much lower (0.04 - 0.1)$\times 10^{22}$ cm$^{-2}$
absorption increasing only to 0.2$\times 10^{22}$ cm$^{-2}$ during the dip.
However, the flares are marked also by a significant increase in the RS intensity. 
 
\begin{figure}
\resizebox{\hsize}{!}{\includegraphics[clip,angle=-90,bb=100 30 560 480]{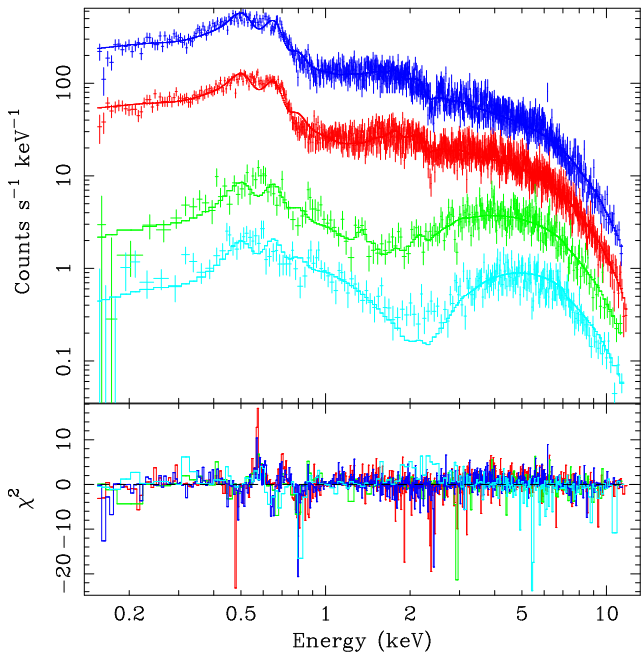}}
\caption[ ]{
The XMM-Newton spectrum of EXO 0748$-$676. Spectra of four different parts
of the orbit
al light curve are shown such as flaring episodes F1 (red) and F2 (blue),
inter-flare Q (green) and dip D (light blue). The spectra were multiplied by
different factors to separate them in the plot (F1 9.0, F2 27.0, 
Q 3.0, D 1.0). The best fitting model (histograms) is a combination of a highly
absorbed power-law and a thin plasma (0.6 keV) thermal emission component with
non-solar abundances and separate (smaller) absorption. The bottom panel shows
the contribution of the residuals to the total $\chi^2$. The spectra are
reproduced by varying mainly the amount of absorption in the power-law
component. Note that the intrinsic power-law intensity is constant, i.e. the
spectra at the high energy end, when not scaled, lie on top of each other}
\end{figure}

\section{Discussion}

The very good sensitivity of the EPIC cameras at energies below 2 keV
allows for the first time to build a consistent picture of the source. A
straightforward interpretation of its variability can be made with an unique
effect of absorption
in front of a stable underlying emission with two separated components.
The hard component, best described by a power-law, is suggestive of a Comptonized-type
spectrum with a high energy break high above the 12 keV XMM-Newton limit.
This same component was also identified in EXOSAT and ASCA observations 
(PA86, Church et al.\ 1998).
However at low energy, the XMM-Newton observations show clearly that the emission is dominated
by a thin hot plasma, affected by a much lower absorption than the hard component. 
This is at variance with previous interpretations involving a second unabsorbed
Comptonized spectrum (PA86) or a blackbody emission (Church et al.\ 1998).

A constraint on the size of the emission area for the hard Comptonized emission 
is provided by
its nearly complete eclipse in the orbital cycle. For a binary orbit
with a P$_{\rm {orb}}$ = 3.82 h, the transition time ($\delta$T) in and out
of eclipses yields a typical size of R = $5.41\times 10^{11}$
(M$_x$+M$_c$)$^{1/3}$$\cdot$($\delta$T/P$_{\rm {orb}}$)$\cdot$sin$i$ cm, where
M$_x$ and M$_c$ are the masses of the compact star and of the companion
in M$_{\sun}$ and $i$ the inclination of the system. Eclipse transition
times are best determined by high time resolution data of 19 eclipses
seen by EXOSAT (PA86) and 10 eclipses seen by RXTE (Hertz et al.\ 1997)
which give mean values of $\sim$~6 s (EXOSAT) and $\sim$~5 s (RXTE).
Assuming a compact star of 1.0 M$_{\sun}$, a 0.45 M$_{\sun}$ companion and
an inclination of $i$ = 75$^{\circ}$ (PA86), yields a typical radius of the
Comptonized emission of $R$ = (2.4$\pm$0.5)$\times 10^{8}$ cm.

On the other hand, the soft thermal component is seen un-eclipsed so that its size
should be significantly greater than the companion occulting sphere which,
for a Roche lobe filling star on the main sequence is of radius 
R $\sim$~7.9$\times 10^{9}$$\cdot$P$_{\rm {orb}}$(hr) cm (Frank et al.\ 1992).
This corresponds to $\sim$~3$\times 10^{10}$ cm for EXO 0748$-$676. 

The picture of the system as seen by XMM-Newton is therefore a rather compact 
($\sim$~2$\times 10^{8}$ cm) Comptonized region as a central hard X-ray source surrounded 
by a more
extended ($\sim$~3$\times 10^{10}$ cm) hot halo-type thermal emission. 
This is the opposite 
to the
previous descriptions in which the central source was expected to 
be a ($\sim$~2\ keV)
blackbody surrounded by a Comptonized emission from the accretion disc corona (ADC)
(Church et al.\ 1998).
We stress that the inner Comptonized region is compatible in size with the 
disturbed inner edge of the accretion disc where low frequency QPOs are expected to 
form. If the observed (0.58 - 2.44) Hz QPOs (Homan et al.\ 1999) are indeed produced 
at the Keplerian frequency in the accretion disc, their distance to the 
centre is given by  $r$~=~1.50~10$^{8}$ $\nu$$^{-2/3}$$\cdot$(M$_x$/M$_{\sun}$) cm, 
where $\nu$ is the QPO frequency, which 
corresponds to a range of (0.8 - 2.2)$\times 10^{8}$ cm in radial distance. The ADC is
then strictly restricted to the central part of the disc.
The spectral variability demonstrates that we always see this inner ADC strongly 
absorbed even during flares, most probably because of the high inclination and
of the occultation by the external parts of the disc including the bulge 
during the dip.

The most extended halo region is much less absorbed. In fact we note that the 
lowest best fitting value of \nh\ is a factor two lower than the average galactic 
column density in this direction \nh\,$\sim$~1.1$\times 10^{21}$ cm$^{-2}$ 
(Dickey \& Lockman 1998). This is possibly due to the more complex shape of the
continuum in this region and a contamination by highly ionized regions in the
system which should be accounted for by a warm absorber. 
The absorption during the dip is a factor $\sim$~2 higher than the galactic value,
revealing that the thermal extended halo is also significantly affected by
absorption of the bulge. This is in favour of a flattened rather than spherical halo.
The emission measure of this thermal halo in quiescent state is 
$\sim$~0.8$\times 10^{58}$~cm$^{-3}$ for a 10 kpc distance. 
For a spherical halo, with a typical dimension of 
$\sim$~3$\times 10^{10}$ cm, the electron density will then be $n$ $\sim$~0.8$\times 10^{13}$~cm$^{-3}$.
If a flat geometry is assumed, with a radius of $\sim$~3$\times 10^{10}$ cm and a typical 
height of $h$ $\sim$~10$^{9}$~cm, this density is significantly higher 
at $n$ $\sim$~5.3$\times 10^{14}$~cm$^{-3}$. 
We note that very similar values of emission measures and densities are derived from
the characteristics of the absorption edges and emission lines of highly ionized
elements clearly detected by the RGS at energies lower than 2.5 keV (Cottam et al.\ 2001).
If the central source is seen through such a flat halo, a rough estimate of the
absorption is \nh\ $\sim$~ $h$.$n$~ $\sim~ 5.3\times 10^{22}$ cm$^{-2}$. This value is consistent
with the measured absorption of 6.4$\times 10^{22}$~cm$^{-2}$ (Table 3).
The estimated dimension and density of the halo suggest that it may be a hot atmosphere
at the surface of an irradiated accretion disc. The high inclination of the system 
will impose then that the central source is always seen through this extended
highly ionized region on top of the accretion disc.
A better description of the halo may therefore be in terms of a fully photo-ionized model which is beyond the scope of this letter.
Some caution should then be taken in interpreting the abundances derived from a
pure collisional RS model.

Assuming a distance of $\sim$~10 kpc, consistent with a maximum Eddington
luminosity in the bursts (Gottwald et al.\ 1986), the (2 - 10 keV) unabsorbed
luminosity of the source is constant at $\sim$~2.0$\times 10^{36}$ erg s$^{-1}$ 
which is comparable to the luminosities observed by ASCA and EXOSAT, 
though at the lowest end. In the total XMM-Newton range, the (0.5 - 10 keV) luminosity only 
slightly varies from 2.64 (in quiescence) to 2.79$\times 10^{36}$ erg s$^{-1}$ (flares).
In this range, the thermal halo luminosity varies from 0.6 (quiescence) to 
2.1$\times 10^{35}$ erg s$^{-1}$ (flares), contributing from 2 to 7.5\% to the total 
luminosity.
The increase of the halo luminosity is apparently not linked to any simultaneous
changes in the central source luminosity. 
The total increase of energy inside the flares with typical duration 
of $\sim$~2000~s is $\sim$~3$\times 10^{38}$ erg. This is only a small fraction of the energy 
$\sim$~2$\times 10^{39}$ erg released during the bursts. It is then possible that the halo 
changing luminosity may be due to a delayed re-processing of a previous burst at
the surface of the accretion disc. 
     \acknowledgements
The results presented are based on observations obtained 
with XMM-Newton, an ESA science mission with instruments and contributions 
directly financed by ESA Member States and the USA (NASA).\\
EPIC was developed by the EPIC Consortium led by the 
Principal Investigator, Dr. M. J. L. Turner. The consortium comprises the
following Institutes: University of Leicester, University of
Birmingham, (UK); CEA/Saclay, IAS Orsay, CESR Toulouse, (France);
IAAP T\"ubingen, MPE Garching,(Germany); IFC Milan, ITESRE Bologna,
IAUP Palermo, Italy. EPIC is funded by: PPARC, CEA, CNES, DLR and ASI.

\end{document}